# Cyber-Security in the Emerging World of 'Smart Everything'

1st Elochukwu A. Ukwandu
*Department of Computer Science,*
*Cardiff School of Technologies,*
*Cardiff Metropolitan University, Cardiff*
Wales, United Kingdom.
eaukwandu@cardiffmet.ac.uk

2nd Ephraim N.C. Okafor
*Department of Electrical and Electronic Engineering,*
*School of Engineering and Engineering Technology*
*Federal University of Technology, Owerri.*
Imo State, Nigeria.
ephraimnc.okafor@futo.edu.ng

3rd Charles Ikerionwu
*Department of Software Engineering,*
*School of Information and Communication Technology*
*Federal University of Technology, Owerri.*
Imo State, Nigeria.
charles.ikerionwu@futo.edu.ng

4th Comfort Olebara
*Department of Computer Science,*
*Faculty of Physical Science*
*Imo State University, Owerri.*
Imo State, Nigeria.
chy_prime@yahoo.com

5th Celestine Ugwu
*Department of Computer Science,*
*Faculty of Physical Science*
*University of Nigeria, Nsukka.*
Enugu State, Nigeria.
celestine.ugwu@unn.edu.ng

Correspondent author: Elochukwu A. Ukwandu *email address: eaukwandu@cardiffmet.ac.uk*

*Abstract*—The fourth industrial revolution (4IR) widely known as Industry 4.0 is a revolution many authors believe have come to stay. It is a revolution that has been fast blurring the line between physical, digital, and biological technologies. These disruptive innovative revolutionary technologies largely rely on high-speed internet connectivity, Cloud technologies, Augmented Reality, Additive Manufacturing, Data science, and Artificial Intelligence. Most developed economies have embraced the revolution while the developing economies are struggling to adopt 4IR because they lack the requisite skills, knowledge, and technology. Thus, this study investigates Nigeria as one of the developing economies to understand her readiness for 4IR and the level of preparedness to mitigate the sophisticated cyber-attacks that comes with the revolution. The investigation adopted quantitative research approach and developed an online questionnaire that was shared amongst the population of interest that includes academic, industry experts, and relevant stakeholders. The questionnaire returned 116 valid responses which were analysed with descriptive statistical tools in SPSS. Results suggest that 60% of the respondents are of the opinion that the Nigerian government at all levels are not showing enough evidence to demonstrate her preparedness to leverage these promised potentials by developing 4IR relevant laws, strong institutional frameworks and policies. They lack significant development capacity to mitigate risks associated with digital ecosystem and cyber ecosystem that are ushered in by the 4IR. In the universities, 52% of the courses offered at the undergraduate and 42% at the post-graduate levels are relevant in the development of skills required in the revolution. The study recommends that the government at all levels make adequate efforts in developing the country's intangible assets such as educational and social capitals as well as strong institutional frameworks, policies and laws that will form the basis for knowledge development, ownership and retention. In all, this paper posits that successful implementation of these could equip Nigeria to embrace the 4IR in all its aspects.

I. INTRODUCTION

The emerging fourth industrial revolution technologies as envisaged in [1], [2], [3], [4], [5] are largely driven by high-speed internet connectivity [2], Cloud technologies, Data science and technologies. It will also rely heavily on Artificial Intelligence (AI) technologies and its sub-fields of Machine Learning, Robotics engineering and technologies, Augmented Reality technology, Additive manufacturing, Smart internetdriven devices such as Internet of Things (IoT), and Industrial Internet of Things (IIoT) devices [6]. The implications of this emerging world as foreseen are likely major disruption in human way of living as day-to-day activities may be predominantly virtual and smart. There are likelihood of continuous reduction of geographical boundaries through video conferencing, tele-medicine, virtual meetings, software driven Virtual learning environment, virtual conferencing, machine-to-machine communication [4], machine-to-human communication [3] and so on. These will in no doubt bring about demand for new skill-sets to help manage evolving nearly and full autonomous systems, artificially intelligent embedded devices, highly connected smart devices, networks and Cloud-dependent technologies.

This also has large implications in Cyber-Security [2] with the emergent of sophistication in cyber-attack landscapes through bio-inspired and Artificial Intelligent (AI)-based cyber-attacks [7]. Furthermore, it will in no small measure affect teaching and learning [8], [9] and hence a demand for content-shift in curriculum and programmes that will support fusion of Cyber-Security and Data Science, Cyber-Ranges and Security Test-Beds, interconnection of Information Technology (IT) to Operational Technology (OT) [4] [3] leading to smart and additive manufacturing, smart everything – city, hospital, transport and so on. In lieu of these dynamism, there will be needs to re-define Privacy for AI efficiency, provide broad AI ethics, practices such as Cyber-hygiene [2] as well as existing laws and regulations. This is because it is projected that in the nearer future, internet may have more information about a person than families and closest relatives should have [2]. The possibilities of new field of studies are expected such as Cyber-Microbiome, Cyber-Anthropology, Cyber-Psychology, Cyber-Data Science and so on.

Beyond recognising these foreseeable implications of the emerging technologies, this study aims to assess how ready academia in developing economies like Nigeria are prepared to leverage the potentials of this high-technologies in this emerging new world so as to leapfrog her social, educational, political development indices. In doing this, the authors of this paper recognises the importance of education as being central to this emerging world by enabling the development of required intangible assets that form the breeding ground for this emerging highly technological and knowledge-driven economy [10]. In view of this, a survey is designed with the aim of eliciting knowledge on the readiness of Nigerian academics of this emerging new world viz-a-viz the conceptualisation, understanding of the implications to teaching and learning as well as research and knowledge development. The efforts in preparing for the requisite classroom administration, manpower needs, existing or emerging legal and policy frameworks to supports the application of these technologies in improving our industries, protect and promote intellectual properties alongside existing infrastructure to support learning and practice of this emerging new way of doing things.

The rest of the paper is organised as follows: Section II focused on critically reviewing relevant literature to ascertain the requisite knowledge gap, while Section III explored existing knowledge on the practises in the industry in relation to cyber-security culture especially within the dynamics of cyber-attack landscape. Section V provides information on how education is being transformed by the emerging world, the requisite skill-sets that will dominate the new world and the views of the stakeholders that providing these new skills will enable the society to benefit from the potentials of the new era. Section VI provided information on the survey materials and method such as population sample and the ways and manner the survey were carried out, while Section VII provided the survey results using Table and Figure for proper visualisation. These were followed with detailed analysis and discussion on the results obtained. The paper concluded with Section VIII, where authors views and findings on the paper were laid with relevant recommendations based on the results.

## II. REVIEW OF RELATED WORKS

Industry 4.0 is premised according to Catal and Tekinerdogan in [11] on emerging new technologies that include Internet of Things, Data Science, Artificial Intelligence and her sub-fields of Deep Learning, Augmented Reality, Edge computing and Digital twins. The authors posit that these bring with it new opportunities with attendant challenges and solutions in many domains of human endeavours. In relation to these disruptive technologies they further stated that requisite set of academic curriculum are needed to tackle them and hence the need to provide frameworks that can support education in the emerging new context of Industry 4.0. The study provided a customised academic courses related to Industry 4.0 for Wageningen University in the Netherlands as a result of their strategic position in World University ranking in Agriculture by applying project-based evaluations. The result of the study as claimed, beside developing academic courses and their curriculum that have paradigm shift in content befitting of Industry 4.0 era, issues like skills in critical thinking, creativity, and problem-solving were addressed by it.

Fitsilis, Tsoutsa and Gerogiannis in their paper in [12] were of the view that due to the perceived disruptive nature of the emerging industry 4.0, the demand to study the competency and knowledge need of the emerging industry has become prominent. Through their study, they presented a competence model to outline the knowledge dimensions and skills of the future, which they classified into technical, behavioural and contextual. Karampidis *et al.* in [13], beyond describing the rationale, aims and objectives of InCyS 4.0 (Industrial CyberSecurity 4.0), outlines the proposed course material for it based on the outcome of their research. InCyS 4.0 is a 2-year education programme targetted at training industrial production technicians in handling cyber-security vulnerabilities in entreprises based in European member nations. The project aims at creating open source educational material that will help address perceived training needs of these class of personnel. The interesting aspect of InCyS 4.0 project according to the authors is that the programme was designed based on empirical evidence from the survey conducted on the security weaknesses of the participating enterprises and the extraction of the profile of the industrial Information Technology engineers that made it adapt the training content to its feedback.

Furthermore, Onar *et al.* in [14] provided a definition of the new education requirements suitable for industry 4.0, they also showed the emerging patterns and similarities in engineering education that will take care of these requirement needs. Mian *et al.* in [15] worked on how university education can be sustained based on different factors capable of influencing the progression and enactment of the system in emerging industry 4.0 era. The results of their study showed some fundamental requirements for universities in Industry 4.0. These include skilled staff, effective financial planning, advanced infrastructure, increased industrial partnerships, revised curricula and carefully tailored insightful workshops.

From the reviews of relevant extant literature, there are commendable interests of the academia in preparing for the emerging disruptively innovative technologies by providing relevant contents, and requisite paradigm shifts in content delivery in the developed economies. Majority of the papers focused in the area of engineering such as InCyS 4.0 programme, but literatures suggest that none has been on Cyber-Security preparedness especially in developing economies. The reason for our interest in developing economies like Nigeria is the perceived potentials of academia to lead in leapfrogging all the indices of social, education and political developments by keying into the emerging disruptive technologies, which is highly knowledge and technologically driven. Furthermore, there are strong indications that Cyber-Security skills will be one of the very demanding skill-sets in the emerging world as the more connected the world becomes by blurring the line between physical, digital and biological technologies using IoT, IIoT devices, embedded systems, etc, the more vulnerable and hence the need for professionals capable of managing the situation.

### III. INDUSTRY 4.0 AND CYBER-SECURITY

Xu *et al.* in [2] opined that industry 4.0 will bring about quite a lot of paradigm shifts in the way life is lived and activities conducted. Far above this, it has appreciable implication in cyber-security as the more connected we become, the more conscious we will be to unethical hackers and cyber-risks as the use of IoT will definitely lead to increase in vulnerabilities in any network and hence calls for greater awareness, preparedness and enhanced skill-sets in cyber-security.

One of the competitive advantage a business can have is the ability to proactively analyse and have fore knowledge of the vulnerability status of main critical assets. These assets need to be protected against cyber-attacks because of their potential business impacts according to Corallo, Lazoi and Lezzi in [16]. The study proposed a structural classification of critical assets of industries within industry 4.0 alongside the potential adverse effects their cyber-security breaches will have on business performance. They also stated that this outcome can be leveraged by academia and industry players in conducting future analysis and investigation in cyber-security.

The challenges industry 4.0 sets to higher education and professional qualification were studied by James and Szymanezyk in [17] using University of Lincoln, United Kingdom as case study. The study proposed the use of what they referred to as Comprehensive, Partial and Merged models of delivery. Thus they posit that the strategy will effectively prepare students with effective relevant industry skills that will make them stand out in Computer Science and Cyber-Security fields.

Threat Sophistication, Landscapes and Industry 4.0: This section reviews prominent cyber-attack landscape as knowing the landscape and dynamics will help predict the likely evolving nature in the emerging industry 4.0. The review will help bring to the fore, the needed knowledge on the evolving attack landscape, predict the future and get prepared through evolving education curriculum, teaching, learning methods, research and knowledge development in Cyber Security in order to leverage the potentials of this disruptive technologies for the good of our society rather than being caught unawares. One of such fast rising attack landscape is in the weaponisation of cyber-attacks using artificial intelligence.

*Cyber-Attack Sophistication:*

- AI-based Attacks The recent advances in AI have been leveraged by cyber-criminals to automate attack processes [18], [19], by taking advantage of technologically enhanced learning, and automation capabilities offered by machine learning and its subsets. The trend has necessitated the pressing need to develop appropriate education curriculum, policies, programmes with focused contents, training methods and technologies.

  Kaloudi and Li in [18] reported a list of existing AI-enhanced cyber-attacks, while Yamin, Ullah and Katt in [20] extensively discussed the extent AI has been used to weaponised cyber-attacks in medicine, traffic management and so on. It is worthy to note that majority of these attacks targeted interconnected and software dependent new generational embedded systems. This is as a result of the integration of information technologies into operational technologies thus narrowing down the line of differentiation between the once two separate technologies and these have wider application in traffic management, healthcare, power, buildings, automotive systems, ships, robots, homes and transport systems.

- Bio-Inspired Attacks Stuxnet attack of 2010 in

Iran [21], [22], [23], [24] is believed by cyber-security stakeholders as an attack that changed the dynamics of cyber-threats. Stuxnet is a malicious computer work that targets supervisory control and data acquisition (SCADA) systems. Because the worm stealthily breach computer system utilising a biological epidemic model in the communication system to propagate the attack, Chen *et al.* [25] coined the term 'A Bio-inspired Transmissive Attack' to describe the scenario as exemplified in Stuxnet. In addition to the hidden nature of the attack, the hacker need not be conversant with the network topology to succeed. Hence, the linkage between transmissive attacks and epidemic models.

In the same vein, de Sa *et al.* in [26] developed the Backtracking Search Optimisation Algorithm (BSA) and Particle Swarm Optimisation (PSO). This is a two Active System Identification attacks developed by using bio-inspired meta-heuristics of Farah, Farah and Farah in [27] and tested it in a controlled environment. The goal was to highlight the potential impacts of automated attacks, especially their degree of accuracy in damaging the Network Controlled Systems, as a stimulus to develop solutions that counter this attack class. In a bid to improve on the accuracy of the estimated models using BSA, de Sa, Carmo and Machadoa in [28] proposed statistical refinement for the outcomes of the two optimization algorithms.

IV. INTANGIBLE ASSETS AND KNOWLEDGE-BASED ECONOMY OF INDUSTRY 4.0

Wealth that cannot be physically quantified in simple terms are referred to as Intangible assets. This is because the wealth in focus here are not physical in nature. The work of Bailey in [10], categorised it into Educational and Social capitals. While educational capital consists of human capital, including raw labour in addition to the sum of knowledge, skills and know-how of the population. Social capital in the other hand, come in form of beliefs, levels of trust, attitudes, behaviours and the quality of formal and informal institutional infrastructure, which includes level of national stability, transparency and other associated elements. Ferreira and Hamilton in [29] supported these views and further posit that the level of trust among people in a society, the efficiency of her judiciary system, clear property rights and effectiveness of government are the yardsticks for measuring intangible assets of a nation. They also went further to strengthen their position by asserting that intangible capital boosts the productivity of labour and results in higher total wealth as well as constitute the largest share of wealth in virtually all countries.

In the same vein, while the rule of law index remains the most suitable way of measuring the level of Social capital of a nation translated as a tool utilised to measure the quality of governance and institutions, and contributes about 57% of rich countries' capital. In the other hand, Education (human capital) is measured through schooling years per capita [30] and provides about 36% of rich nation's intangible assets [10], [29]. According to Cobb in [30] studies have shown that nation's with well-educated population as well as stable investment-friendly environment have the potentials of producing greater national wealth as against that with under-educated and unstable societies.

Education, according to Ilori and Ajagunna in [31] especially higher education with requisite attention and investments has been source of innovation, policy and legal frameworks, knowledge development and a great national asset through empirical studies, knowledge creation and dissemination using appropriate tools. These in turn strengthens the position by the authors of this paper that one of the major ways digital economy as driven by knowledge can be maximised are when efforts of government and relevant national stakeholders are geared towards providing strong national institutions, quality and accessible education to the citizens. As these assure the promotion and protection of intellectual properties, such as patents, trademarks and copyrights as well as brand recognition that are major drivers of the digital eco-system around the world. This position is supported by the works of Gancia and Zilibotti in [32], Chen, Niebel and Saam in [25], Cardona, Kretschmer and Strobel in [33] and Kretschmer in [34].

V. EDUCATION CURRICULUM AND INDUSTRY 4.0

Education, especially higher education with requisite attention and investments has been source of innovation, policy, knowledge development and a great national asset [31]. According to Xu *et al* in [2], the potentials of Industry 4.0 is disruptively innovative with likely impacts on core industries and sectors like education, healthcare and business. Within the context of industry 4.0, education as a service industry will shift its focus from a customer learning model of the third industrial revolution. This is largely because of the blurring technological lines between physical, digital and biological technologies as shown in [2], [3] and hence will change the way of content delivery. This change will have impact in educational curriculum, teaching techniques and mode of deliveries with focus changing from mode of teaching to mode of learning. In all, teaching and learning model will be predominantly blended in approach by combining the traditional face-to-face teaching and learning model with virtual teaching and learning model using software driven virtual learning environment that is delivered and administered over the internet, which the authors refer to as 'phy-digital' teaching and learning environment.

For this emerging non-reversible revolution [35] to be largely positive in such a manner as to benefit our social, environmental, economic and political lives will depend largely on the efforts of the stakeholders - researchers, educators, developers and regulators in keying into this proactively [3]. This anticipated collaboration will be geared towards developing understanding and skills commiserate and fit enough to help navigate an increasingly convergent complex unfolding landscape of industry 4.0 through an overhaul of educational curriculum, teaching and learning methods, administrative technologies, legal and policy frameworks, value systems and cultural landscapes.

## VI. METHOD AND SURVEY MATERIALS

This paper aims to ascertain the information gap in the area of teaching and learning, skill development, policy and legal frameworks in Cyber Security with regards to the emerging industry 4.0 in a developing economy, such as Nigeria. The research design followed a quantitative approach and administered an online questionnaire to the academic, industry players, researchers and requisite stakeholders to elicit the level of awareness, preparedness and skills been bequeathed in relation to the emerging disruptive industry 4.0 era. The survey focuses on investigating the level of awareness, understanding, preparedness through projected curriculum, technologies, and administration in Nigerian universities. Additionally, it investigated the research interests and direction being propagated by the relevant stakeholders to help in the dissemination of knowledge, requisite skill-sets. It also extends to the development of the policy and legal frameworks that would be used in mitigating this emerging dynamic cyber-threat landscapes as posited by Xu *et al.* in [2].

The questionnaire considered 21 questions bordering on the main objective of this study. Probable respondents were identified from the industry and academia ranging from Union leaders, Head of Departments, Directors, Senior and Middle-level career personnel. Although the scope of the study is limited to Nigeria, seven Technology Universities, one premier university, two conventional universities and an Agency in the Ministry of Communication and Digital Economy were selected because of their respective areas of specialisations in science, technology, and engineering. The Agency was selected as a result of her mandate in human capacity development both in the academia, government Ministries, Departments and Agencies in the areas of computer literacy, digital soft skills and economic development. The questionnaire was distributed to the industry experts through their emails, which were obtained from their websites and several inquiries made through their respective customer service officers. Similarly, contacts where established in each of these universities and through them, the questionnaire was shared through their WhatsApp groups.

## VII. ANALYSIS, RESULTS AND DISCUSSIONS

In order to achieve the main objective of this study, two research questions were considered:

1) To what extent is Nigeria prepared in leveraging the emerging fourth industrial revolution to leapfrog her requisite development indices.
2) To what extent of readiness are relevant industries prepared to mitigate the attendant cyber-attack sophistication that comes with high level network connectivity as expected in fourth industrial revolution of nearly 'smart everything'

After two months of sharing the questionnaire, a total of 116 valid responses were received. Statistical tools in SPSS were used to perform analysis on the data received using Descriptive statistics. Tables I-VIII present the results from the valid responses from the questionnaire and indicates the number of participants from the sample size of 116 that have:

- Result 1: knowledge (aware) of the emerging 4th industrial revolution
- Result 2: conducted 4IR compliant research in their respective fields such as an application of Machine learning in automating a process for enhanced outputs
- Result 3: taught or teaching 4IR compliant subjects such as Additive manufacturing, Cyber Security, etc. Above that, the number of participants in the academic institutions that have
- Result 4: attended training in any of the emerging new fields. Furthermore, the number of:
- Result 5: first degree and graduate degree programmes that are 4IR relevant being offered in their respective institutions. Finally on their assessment of:
- Result 6: Nigerian universities, industries and government institutions - Ministries, Departments and Agencies readiness for the emerging new world.

*A. Result 1: Respondents with knowledge (aware) of the emerging 4th industrial revolution.*

Table I
THOSE AWARE OF EMERGING 4IR

| Valid | Frequency | Percent | Valid Percent | Cumulative Percent |
|---|---|---|---|---|
| No | 45 | 38.5 | 39.1 | 39.1 |
| Yes | 70 | 59.8 | 60.9 | 100 |
| Total | 116 | 98.3 | 100 | |

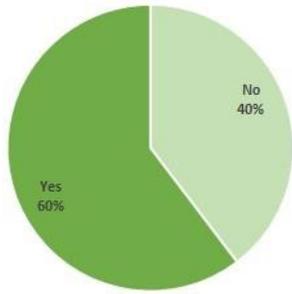

Figure 1. Number of participants aware of emerging 4IR

Results from Table I and Figure 1 show that 60% of the respondents have knowledge (aware) of the concept - 4th industrial revolution. This therefore provides room for further interrogations on what they understand, their views on what the concept entails and the levels of preparedness envisaged in both University, industry and the nation in general.

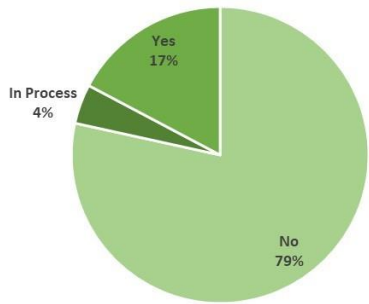

Table II
NUMBER OF INSTITUTIONAL BASED RESEARCH RELEVANT TO 4IR

| Valid | Frequency | Percent | Valid Percent | Cumulative Percent |
|---|---|---|---|---|
| No | 91 | 79 | 79 | 79 |
| In Process | 5 | 4.3 | 4.3 | 83.3 |
| Yes | 5 | 16.7 | 16.7 | 100 |
| Total | 116 | 100 | 100 | |

Figure 2. Number of participants that have conducted 4IR relevant research

B.     Result 2: The number of respondents that have conducted 4IR compliant research in their respective fields such as an application of Machine learning in automating a process for enhanced outputs.

From Table II with visualisation in Figure 2, 79% of the respondents have not conducted any 4IR related research such as in the areas of Machine Learning, Robotics, Addictive manufacturing, Augmented Reality and so on. This implies that despite having greater percentage of the respondents agreeing to know about the concept, there is no significant contribution from the academic world in providing empirical results that will guide in developing appropriate policies in line with the requirements of the emerging 4th industrial revolution. Since 17% of the respondents agreed to have conducted such relevant research, their contributions are not to be deemed sufficient and hence conclude that not much are being done in developing requisite knowledge through empirical studies in these domains in Nigeria.

C.     Result 3: The number of respondents that have are teaching or may have taught 4IR compliant subjects such as Additive manufacturing, Cyber Security, etc

Table III
TEACHING 4IR RELEVANT SUBJECTS

| Valid | Frequency | Percent | Valid Percent | Cumulative Percent |
|---|---|---|---|---|
| No | 84 | 72.4 | 72.4 | 72.4 |
| Yes | 32 | 27.6 | 27.6 | 100 |
| Total | 116 | 100 | 100 | |

Figure 3. Number of participants that are teaching 4IR relevant courses

Table IV
WHEN ATTENDED A 4IR RELEVANT TRAINING

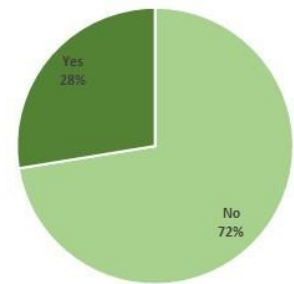

| Valid | Frequency | Percent | Valid Percent | Cumulative Percent |
|---|---|---|---|---|
| None | 41 | 35.3 | 35.3 | 35.3 |
| <1yr | 37 | 31.9 | 31.9 | 67.2 |
| 1-2yrs | 23 | 19.8 | 19.8 | 87.0 |
| 3-4yrs | 15 | 13.0 | 13.0 | 100 |
| Total | 116 | 100 | 100 | |

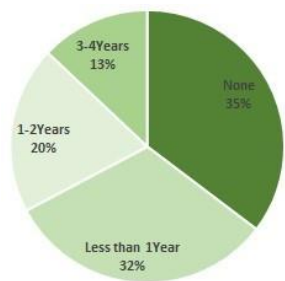

Figure 4. The last time participants attended 4IR relevant training

### D. Result 4: The number of respondents that have attended training in any of the emerging new fields in the last couple of years.

Table III and Figure 3 from the survey results show that only 28% of the academics agree they are teaching 4IR relevant subjects in their various institutions of higher learning. While 72% are of the opinion that they are not teaching anything relevant to the emerging world. These results suggest that relevant educational stakeholders in Nigeria appear not to be doing much with regards to developing requisite subjects or courses and curriculum tailored in manpower training in the emerging new technologies. The implications of these are enormous and cannot be over-emphasised as not having requisite manpower in the emerging domains of human endeavours will limit the opportunities of the citizens and the nation in leveraging the potentials offered by these smart technologies.

However, with about 37% of the respondents in Table IV as visualised in Figure 4 agreeing to have attended training relevant to the emerging world in recent times in comparison with 20% obtained in the last 2-3 years and 13% in the 3-4 years. These show an increasing awareness, training and activities in developing the requisite skill-sets and by extension manpower required to fill in the gap in the coming years in Nigeria.

### E. Result 5: The number of degree programmes and graduate degrees that are 4IR relevant being offered in their respective institutions.

Table V
NUMBER OF DEGREE PROGRAMME THAT ARE 4IR RELEVANT IN THEIR RESPECTIVE UNIVERSITIES

| Valid | Frequency | Percent | Valid Percent | Cumulative Percent |
|---|---|---|---|---|
| None | 48 | 41.4 | 41.4 | 41.4 |
| 1-2 | 33 | 28.4 | 28.4 | 69.8 |
| 3-4 | 14 | 12.1 | 12.1 | 81.9 |
| 5+ | 21 | 18.1 | 18.1 | 100 |
| Total | 116 | 100 | 100 | |

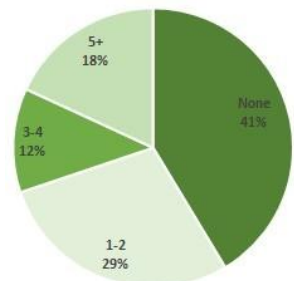

Figure 5. Number of 4IR relevant degree programme in their respective Universities

Table VI
NUMBER OF GRADUATE DEGREE PROGRAMMES THAT ARE 4IR RELEVANT IN THEIR RESPECTIVE UNIVERSITIES

| Valid | Frequency | Percent | Valid Percent | Cumulative Percent |
|---|---|---|---|---|
| None | 53 | 45.7 | 45.7 | 45.7 |
| 1-2 | 32 | 27.6 | 27.6 | 73.3 |
| 3-4 | 14 | 12.0 | 12.0 | 85.3 |
| 5+ | 17 | 14.7 | 14.7 | 100 |
| Total | 116 | 100 | 100 | |

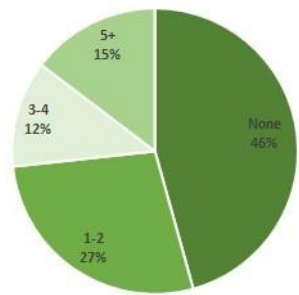

Figure 6. Number of 4IR relevant graduate degree programmes in their respective Universities

With 59% of the respondents in Table V and Figure 5 agreeing that their various institutions offer varying undergraduate degree programmes in 4IR relevant areas; as 29% confirmed offering 1-2 undergraduate programmes, while 12% offer about 3-4 of such programmes and 18% of from 5 programmes and above. This percentage surpasses the 41% of the respondents who said their institutions are not in any way offering any 4IR compliant undergraduate programmes.

The above results are similar to the responses received with regards to graduate programmes undertaken by various institutions of higher learning in Nigeria. From the responses, 54% as shown in Table VI and Figure 6 offer various 4IR compliant graduate degree programmes. Of these responses 27% posit that their institutions have about 1-2 graduate degrees being awarded in 4IR relevant fields, while about 12% consent to having 3-4 of such programmes and 15% shows the availability of 5 and above of such graduate degrees.

Comparing the results from Tables V, VI and Figures 5, 6 with that of Tables II, III and Figures 2, 3 provide good insight for further analysis. The elements of first degree or graduates degree programmes any where in the world are the subjects or courses they contain. That is to say, each programme is made of subjects or courses with curriculum tailored to meet its aim and objectives in manpower development through impacting requisite skill-sets in line with the discipline's core mandates. A closer look at responses from Table III as elaborated in Figure 3 show that the institutions represented are not doing much in teaching requisite subjects or courses that are 4IR compliant. It therefore suggests that if a good number of programmes in 4IR relevant fields exist without having requisite subjects or courses being taught in these fields or disciplines, the programmes being undertaken in these institutions may be deemed as more of buzz words or nomenclatures.

*F. Result 6: Measuring Nigerian universities, industries and government institutions readiness for the emerging new world through the opinions of the major stakeholders.*

Table VII
VIEWS ON UNIVERSITIES READINESS FOR 4IR COMPLIANCE

| Valid | Frequency | Percent | Valid Percent | Cumulative Percent |
|---|---|---|---|---|
| No | 70 | 60.3 | 60.3 | 60.3 |
| Maybe | 17 | 14.7 | 14.7 | 75.0 |
| Yes | 29 | 25.0 | 25.0 | 100 |
| Total | 116 | 100 | 100 | |

With about 60% of the respondents in Table VII, Figure 7 and 60% of the respondents in Table VIII and Figure 8 indicating that higher educational institutions in Nigeria, industries and government are not showing strong commitment in developing programmes, policies and laws that will enable her leverage the advantages of the emerging fields in closing their development gaps. It is therefore the opinion of the authors that the above results and discussions provide empirical evidence to answer these research questions:
- how prepared Nigeria is in leveraging the emerging fourth industrial revolution to leapfrog her requisite development indices? and
- how prepared the relevant industries are in mitigating the attendant cyber-attack sophistication that comes with high level network connectivity as expected in fourth industrial revolution of nearly 'smart everything'.

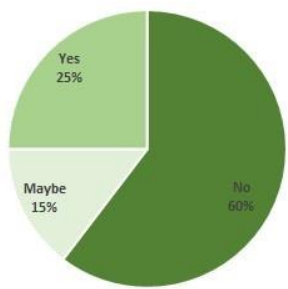

Figure 7. Figure showing views on Universities' readiness for 4IR compliance

Table VIII
VIEWS ON NIGERIAN'S READINESS FOR 4IR COMPLIANCE

| Valid | Frequency | Percent | Valid Percent | Cumulative Percent |
|---|---|---|---|---|
| No | 69 | 59.5 | 59.5 | 59.5 |
| Maybe | 26 | 22.5 | 22.5 | 82 |
| Yes | 21 | 18.0 | 18.0 | 100 |
| Total | 116 | 100 | 100 | |

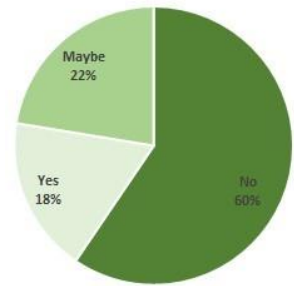

Figure 8. Figure showing views on Nigerian's readiness for 4IR compliance

## VIII. CONCLUSION AND RECOMMENDATIONS

The emerging fourth industrial revolution has been predicted to be very disruptive in ways and manner human conduct daily life activities. It has also been predicted to be a mixture of virtual and real-life activities with modern machine-to-machine communications, which will largely reduce human interventions in production industry and service deliveries. It is in line with the foregoing, this research was premised. It focuses on finding out how prepared developing economies like Nigeria is in tapping into the promised potentials of the budding new world so as to create better living conditions to her largely youthful population. It went further to examine the level of preparedness in mitigating the foreseeable proliferation of threats, especially computer-based threats such as cyber-attacks that come with highly connected computer systems, networks and devices.

This work was able to answer two research questions on the level of Nigerians preparedness in tapping into the promises of 4IR to better the living conditions of her citizens as well as that of capacity development in standing up against the attendant risks. From the survey conducted, 60% of the respondents are of the view that Nigerian governments at all levels are not showing enough evidence to demonstrate her preparedness to leverage these promised potentials as well as making adequate efforts in developing capacity to mitigate risks associated with digital eco-system.

In line with developing requisite skill-sets commensurate with the emerging technologies by the higher educational institutions, available results suggest that a good number programmes in 4IR relevant fields are being offered in both undergraduate (52%) and post-graduate (47%) degree levels. But in the contrary there are few relevant and compulsory courses (32%) in these areas being taught in the respective disciplines in these institutions, which is casting doubt about the efficacy of these programmes and the ability to produce relevant skill-sets that will match the emerging technological know-hows as well as meet various aim and and objectives of these disciplines in manpower development. However, responses suggest that, there are increasing awareness and training over the last five years by the academic stakeholders in these directions.

Thus, this paper recommends that these tempo on training be sustained. Moreso, the academic community should focus more attention in conducting deep rooted scientific research around the emerging new field as a way of developing their own capacities to impact relevant skill-sets and as a way of influencing policy directions, strengthening the development of legal and policy frameworks needed to protect intellectual properties, copyrights, and conducive environment. This would encourage the development of intangible assets required to spur high skilled knowledge development required in the emerging disruptive world.

ACKNOWLEDGEMENTS

Not applicable.

FUNDING

Not applicable.

ABBREVIATIONS   All acronyms were properly defined in the paper.

AVAILABILITY OF DATA AND MATERIALS

All data analysed and used in this paper are both secondary and primary, while the secondary data were publicly available data, the primary were collected using an online questionnaire administered to selected respondents.

COMPETING INTERESTS

The authors declare that they have no competing interests.

CONSENT FOR PUBLICATION

All authors have read and agreed to the published version of the manuscript.

AUTHORS' CONTRIBUTIONS

This paper was conceptualised and formally analysed by Elochukwu Ukwandu and Ephraim Okafor. While Elochukwu Ukwandu and Charles Ikerionwu carried out the investigation. The methodology was developed by Elochukwu Ukwandu and Comfort Olebara. Furthermore, the administration and supervision of this paper were carried out by Elochukwu Ukwandu, Ephraim Okafor, Celestine Ugwu, Comfort Olebara, and Charles Ikerionwu. Comfort Olebara did the data analysis and Celestine Ugwu did the result validation. Formal writing of the paper that produced original draft were done by Elochukwu Ukwandu and Charles Ikerionwu. Elochukwu Ukwandu, Charles Ikerionwu, Ephraim Okafor, Comfort Olebara and Celestine Ugwu reviewed and edited the paper. All authors have read and agreed to the published version of the manuscript.